# Chromium Self-Traceable Length Standard: Investigating Geometry and Diffraction for Length Traceability Chain

Zichao Lin[1,2,3,4,5], Yulin Yao[1,2,3,4,5], Zhangning Xie[1,2,3,4,5], Dongbai Xue[1,2,3,4,5], Tong Zhou[1,2,3,4,5], Zhaohui Tang[1,2,3,4,5], Lihua Lei[6], Tao Jin[7], Xiong Dun[1,2,3,4,5], Xiao Deng[1,2,3,4,5]*, Xinbin Cheng[1,2,3,4,5], Tongbao Li[1,2,3,4,5]

1. Institute of Precision Optical Engineering, Tongji University, Shanghai 200092, China;
2. MOE Key Laboratory of Advanced Micro-Structured Materials, Tongji University, Shanghai 200092, China;
3. Shanghai Frontiers Science Center of Digital Optics, Tongji University, Shanghai 200092, China;
4. Shanghai Professional Technical Service Platform for Full-Spectrum and High-Performance Optical Thin Film Devices and Applications, Tongji University, Shanghai 200092, China;
5. School of Physics Science and Engineering, Tongji University, Shanghai 200092, China;
6. Shanghai Institute of Measurement and Testing Technology, Shanghai 201203, China;
7. University of Shanghai for Science and Technology, Shanghai 200093, China
* Correspondence: 18135@tongji.edu.cn

**Abstract:**
Natural constant-based metrology methods offer an effective approach to achieving traceability in nanometric measurements. The Cr grating, fabricated by atom lithography and featuring a pitch of $d$=212.7705±0.0049 nm traceable to the Cr transition frequency $^7S_3 \rightarrow {}^7P_4^0$, demonstrates potential as a self-traceable length standard in nano-length metrology by grating interferometer. This research aims to analyze and engineer the diffraction characteristics that enhance the Cr grating as a self-traceable length standard within the length traceability chain based on the Cr transition frequency. Accordingly, we investigate the geometric morphology and diffraction characteristics of the Cr grating, analyzes the influence of the grating's polarization-sensitive characteristics on the Littrow configuration grating interferometer, and establishes the criteria for Cr grating fabrication. Experimentally, we fabricate an expanded Cr grating by scanning atom lithography, characterize its diffraction performance, and conduct preliminary verification of length measurement in a self-traceable grating interferometer. This work adheres to the international trend of flattened metrology development, offering a valuable reference for advancing subsequent metrological technologies throughout the new traceability chain.

## 1. Introduction

In 2019, the International System of Units (SI) redefined all basic units using natural constants, reflecting the trend in metrology towards traceability through natural constants[1-3]. In nano-length metrology, laser interferometers are the primary method for transferring SI "meters" in the measurement and calibration of nano-positioning and microstructure dimensions due to their direct traceability, high accuracy, and non-contact characteristics[4-6]. However, limitations related to the refractive index of air [7, 8], have led to an increasing interest in substituting laser wavelength standards with physical period standards [9]. The silicon lattice parameter ($d_{220}$=192.0155714×10$^{-12}$ m, with a physical standard uncertainty of 0.0000032×10$^{-12}$ m), enables a native 0.192 nm resolution using an X-ray interferometer without air refractive index calibration and electronic subdivision[10-13]. Since 2019, it has been recommended by the BIPM as a secondary reproduction of the SI definition of "meter" [11,14]. Although its high accuracy, practical challenges such as single-crystal silicon fabrication, stringent light source requirements (Cu Kα rays = 0.154 nm), slow measurement speeds (10 μm s$^{-1}$), and limited measurement ranges (~10 μm) restrict its widespread industrial adoption[15].

The Cr grating fabricated by atom lithography (also referred to as laser-focused atomic deposition[16,

17]) share the self-traceability feature with Si. In Fig.1, the "self-traceability" concept signifies that the pitch of the Cr grating can be traceable to the natural constant of Cr transition frequency $^7S_3 \rightarrow\ ^7P_4^0$ ($\lambda$=425.55 nm), ensuring pitch uncertainty at the picometer level ($d = \lambda/2$=212.7705±0.0049 nm) [18]. This suggests that Cr grating possess metrological significance comparable to Si and can similarly enable the direct traceability of the SI definition of "meter" through self-traceable grating interferometer (reference standard: Cr grating pitch $d$). Moreover, unlike the Si process that relies on X-ray sources, the Cr grating approach benefits from utilizing readily available light sources with $\lambda < 2d$ = 425.6 nm, such as 405 nm or 375 nm. Concurrently, our group has successfully pioneered the transition from μm-scale to mm-scale fabrication of Cr grating[19], and implemented a self-traceable displacement measurement system with >1 mm/s speed, a native $d/2$=106.39 nm resolution and sub-nanometer accuracy[20].

Therefore, with the advancements in grating fabrication and the development of a self-traceable displacement measurement system, the application of Cr grating is no longer limited to the traditional metrology calibration of metrological instruments such as AFM and SEM[21-25]. To facilitate the direct utilization of natural constants in nanotechnology, we are currently establishing a length traceability chain based on Cr transition frequency[20]. This traceability chain, which employs diffraction light interference to disseminate and ensure traceability for the self-traceable length standard of Cr grating, starts with the SI "Meter" and creates a "Cr grating-212.77 nm pitch standard" by atom lithography. Subsequently, a Littrow-configured "Self-traceable grating interferometer" is developed for "Precise positioning application". Previously, there has been insufficient in-depth understanding of Cr grating as sub-wavelength metallic gratings, specifically regarding their geometric morphology and diffraction characteristic distributions. This lack of detailed understanding has hindered the optimization of self-traceable grating interferometer in precise positioning applications. Addressing this critical gap, this research aims to conduct a comprehensive analysis of the optical diffraction characteristics of Cr grating, with the criteria of establishing fabrication direction that enhance the Cr grating as a self-traceable length standard within the length traceability chain.

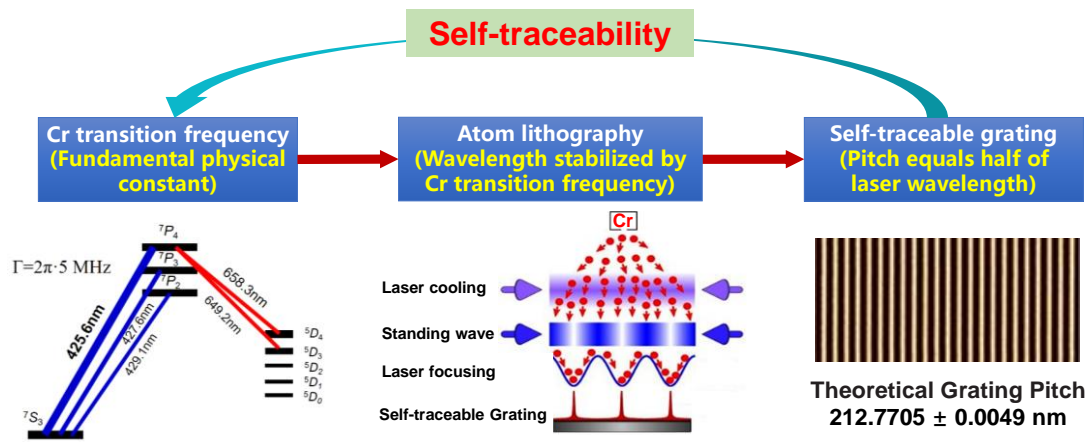

**Fig. 1.** Self-traceability of Cr grating fabricated by atom lithography[16,17]

## 2. Theoretical explanation

*2.1 Definition of Cr grating grooves*

*2.1.1 Gaussian-symmetric chromium atom lithography grating*

In atom lithography, the longitudinal height and transverse width of the Cr grating grooves, which determine the typical Gaussian distribution characteristics, are directly influenced by the deposition thickness and surface growth of the atoms[26-28].

Thus, the grating groove function $f(x)$ depicted in Fig. 1 is defined as follows:

$$f(x) = H_0 \exp\left(-\frac{x^2}{2c^2}\right), x \in \left[-\frac{d}{2}, \frac{d}{2}\right] \tag{1}$$

where $H_0$ and $F_0 = 2\sqrt{2\ln 2}c$ represent the height and width of the Gaussian function, respectively; $d$ is the grating pitch.

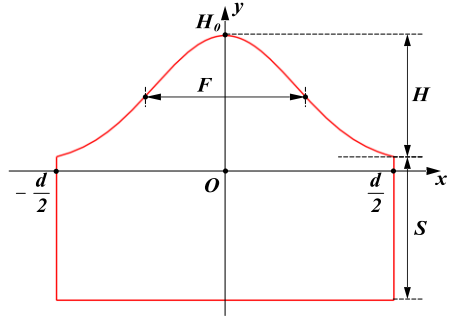

**Fig. 2.** Symmetric groove definition of the Cr grating, where $H$ is the peak-to-valley height, and $F$ is the full width at half maximum. The grating pitch $d$ is precisely 212.78 nm and is traceable to the Cr transition frequency from $^7S_3 \to {}^7P_4^0$.

The Cr grating grooves can be characterized by three geometric parameters: $H$, the peak-to-valley height (PTVH); $F$, the full width at half maximum (FWHM); and $S$, the valley-to-substrate height (VTSH). The contrast $C$, defined as $C=H/(H+S)$, describes the rate of chromium atom conversion during the deposition process to form the grating structure. In this work, the value is set at $C=0.5$, which satisfying the condition $H=S$ [29]. Due to the grating pitch $d$ being used as the independent variable interval in $f(x)$, $H$ and $F$ are related as follows:

$$H = H_0\left(1 - \exp\left(-\frac{d^2}{4c^2}\right)\right) \tag{2}$$

$$F_0^2 = -\frac{F^2}{\ln 2} \times \ln\left(\frac{1}{2}\left(1 + 2^{-\frac{d^2}{F^2}}\right)\right) \tag{3}$$

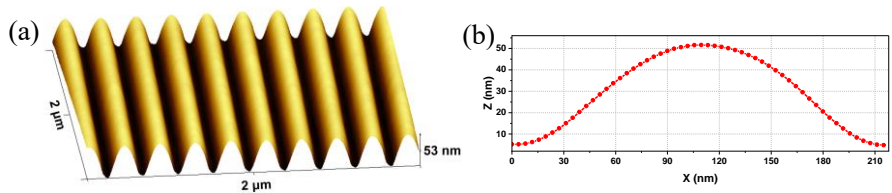

**Fig. 3.** (a) AFM 3D scan image of Cr grating with $H$ = 53 nm, $F$ = 118 nm, $d$ = 212.78 nm. (b) Magnified AFM image of the grating groove, showing a Gaussian distribution.

Fig. 3(a) displays the AFM 3D scan result of the Cr grating, while Fig. 3(b) presents a local zoom with a Gaussian distribution. The Gaussian shape of the grating grooves stems from two primary mechanisms. Lateral structure expansion due to the atoms surface growth in the Cr grating, as reflected by parameter $F$, results in the widening of the feature width FWHM. Meanwhile, parameter $H$ captures the variation in the vertical structure, which is driven by differences in Cr atoms flux and deposition time, inducing changes in PTVH. These two parameters indicate that the geometrical degrees of freedom for the Cr grating are finite (primarily controlled by $H$ and $F$). This limitation imposes challenges the optical design for the diffraction characteristics of the Cr grating.

*2.1.2 Non-Gaussian symmetric chromium atom lithography grating*

In Fig. 4, the atom lithography deposition conditions, which are affected by inhomogeneous standing wave fields, inadequate collimation of the atomic beam, and diffraction at the substrate edges[30], often result in a central offset of the grating structure with a value of $s$. However, considering that the grating structure is essentially a statistical result of the convergence of a large number of Cr atoms toward the wave node, influenced by the standing wave field. Consequently, the Gaussian distribution characteristics of the groove remain unchanged.

Thus, $f(x)$ can be reconstructed as follows:

$$f'(x) = \begin{cases} H_0 \exp\left(-\frac{(x/k_L)^2}{2c^2}\right), & x \in \left[-\frac{d}{2}-s, 0\right] \\ H_0 \exp\left(-\frac{(x/k_R)^2}{2c^2}\right), & x \in \left[0, \frac{d}{2}-s\right] \end{cases} \quad (4)$$

where $k_L = 1 + \frac{s}{d/2}$ $k_L$=1+2s/d, $k_R = 1 - \frac{s}{d/2}$ and $s$ is the scaling factor of the $f'(x)$.

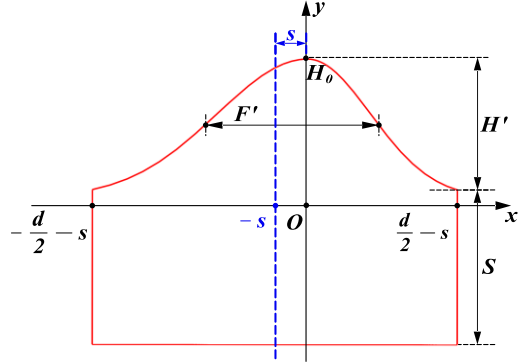

**Fig. 4.** Asymmetric groove definition of Cr grating, $s$ is the center offset of the grating. The groove characteristics approximately maintain the relationship $H' = H$, $F' = F$.

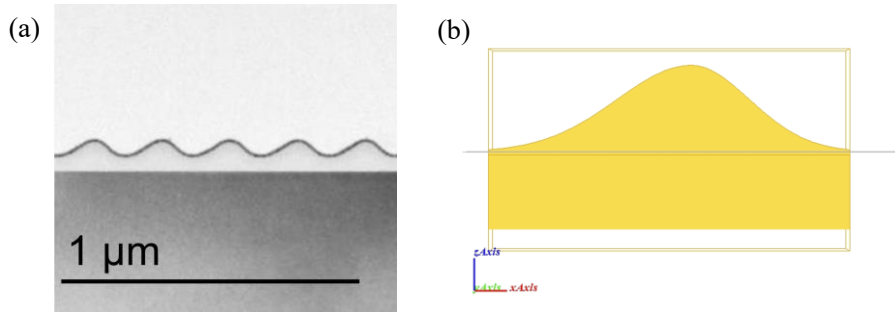

**Fig. 5.** (a) Asymmetric TEM image of the Cr grating. (b) Schematic diagram of the cross section of the asymmetric model, constructed using Gallop EM simulation software, where $H$=50 nm, $F$=90 nm, $s$=15 nm.

Fig. 5(a) displays a TEM image of the asymmetric Cr grating, which demonstrates an overall shift of the grating structure to the right side due to unfavorable experimental deposition conditions. Fig. 5(b) illustrates the corresponding schematic diagram of the model constructed using $f'(x)$, which shows a high degree of consistency with the TEM image. In conclusion, the two Gaussian groove equations provided in Eq. (1) and (4) can effectively describe various Cr grating preparation outcomes and act as the basis for subsequent diffraction property simulations.

*2.2. Theoretical analysis*

In this section, we examine the polarization sensitivity of Cr gratings (4700 l/mm) as sub-wavelength metallic gratings in the self-traceable grating interferometer under a 405nm light source. Our objective is to define diffraction characteristics criteria for the Cr grating, optimizing its design to improve performance in the measurement system and facilitate the development of the length traceability chain.

*2.2.1 Self-traceable grating interferometer setup*

Fig. 6 depicts the optical path schematic of the Littrow-configured "self-traceable grating interferometer". A more detailed explanation of the Littrow-configured optical path can be found in [31],The paths of the two measurement beams can be described as follows:

measurement beams 1:Laser-PBS-QWP$_1$-M$_1$-Cr Grating-M$_1$-QWP$_1$-PBS

measurement beams 2:Laser-PBS-QWP$_2$-M$_2$-Cr Grating-M$_2$-QWP$_2$-PBS

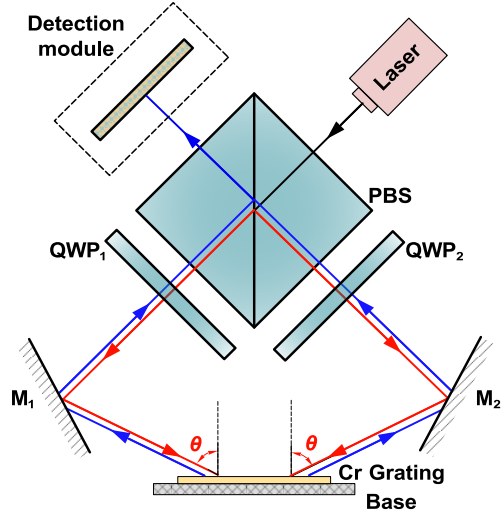

**Fig. 6.** Littrow configuration schematic of the self-traceable grating interferometer.

Setting $A_0$ represent the amplitudes of the s-polarized light and p-polarized light. $A_+$ and $A_-$ represent the amplitudes of the measured beams entering the detection module. The energy conversion efficiency $\eta_{\pm L}$ of the Littrow-configuration grating interferometer (LCGI) can be defined as:

$$\eta_{\pm L} = \left(\frac{A_\pm}{A_0}\right)^2 \qquad (5)$$

When the Cr grating shifts a surface distance $\Delta x$, the ensuing alterations in the interference light intensity $I$ and contrast $K$ are given by Eq. (6) and Eq. (7). It's important to note that the measurement of the interference light intensity $I$ is anchored to the Cr grating period $d$, which is traceable to the Cr transition frequency $^7S_3 \rightarrow {}^7P_4^0$. These principles form the foundation of the displacement-to-interference mapping for a self-traceable grating interferometer:

$$I \propto A_0^2 \left( \eta_{+L} + \eta_{-L} + 2\sqrt{\eta_{+L}\eta_{-L}} \cos\left(\frac{4\pi}{d}\Delta x\right) \right) \tag{6}$$

$$K = \frac{I_{max} - I_{min}}{I_{max} + I_{min}} = \frac{2\sqrt{\eta_{+L}\eta_{-L}}}{\eta_{+L} + \eta_{-L}} \tag{7}$$

Achieving high-accuracy demodulation of the displacement $\Delta x$ requires a high-quality interferometric signal, with a high signal amplitude playing a key role in improving the signal-to-noise ratio[32]. According to Eq. (7), this requires that:

$$\eta_{+L} = \eta_{-L} = \eta_{L\text{-max}} \tag{8}$$

where $\eta_{L\text{-max}}$ is the potential maximum energy conversion efficiency of LCGI, which is directly related to the grating diffraction characteristics. The subsequent section is dedicated to the exploration of optimizing the performance of LCGI by modifying the diffraction characteristics of the grating.

*2.2.2 Fabrication criteria of grating diffraction characteristics in LCGI*

Given the ideal optical components in Fig. 6, consider the grating's Jones matrix and its modulation effect on the complex amplitude of incident light:

$$G = \begin{bmatrix} -\sqrt{\eta_{TM}} & 0 \\ 0 & \sqrt{\eta_{TE}} \exp(i\Delta\delta) \end{bmatrix} \exp\left(i\frac{4\pi}{d}\Delta x\right) \tag{9}$$

where $\eta_{TM}$ and $\eta_{TE}$ are the diffraction efficiency of the grating in the TM and TE directions, respectively. $\Delta\delta$ is the phase difference between the diffracted light in the two directions, which takes the value of $(-\pi, \pi]$.

Considering Eq. (9), when $\eta_{TM} \neq \eta_{TE}$ or $\Delta\delta \neq 0$, residual polarization originating from diffracted light at the PBS can leak back to the laser. This leakage reduces the system's energy conversion efficiency and causes fluctuations in the laser output power[33]. Furthermore, a lower -1st order diffraction efficiency can lead to a higher electrical noise percentage in the interference signal, consequently affecting displacement measurement accuracy. These observations emphasize the crucial role of the grating's diffraction characteristics in determining interference performance.

Applying Eq. (9) to Fig. 6, the measurement beams that finally enter the detection module can be expressed as:

$$E_+ = PBS_s \cdot QWP_{1(-\theta)} \cdot M_1 \cdot G \cdot M_1 \cdot QWP_{1(\theta)} \cdot PBS_p \cdot E_0 \tag{10}$$

$$E_- = PBS_p \cdot QWP_{2(-\theta)} \cdot M_2 \cdot G \cdot M_2 \cdot QWP_{2(\theta)} \cdot PBS_s \cdot E_0 \tag{11}$$

where $PBS_p = \begin{bmatrix} 1 & 0 \\ 0 & 0 \end{bmatrix}$, $PBS_s = \begin{bmatrix} 0 & 0 \\ 0 & -1 \end{bmatrix}$, $M = \begin{bmatrix} -1 & 0 \\ 0 & 1 \end{bmatrix}$, $QWP_\theta = \begin{bmatrix} 1 - i\cos 2\theta & -i\sin 2\theta \\ -i\sin 2\theta & 1 + i\cos 2\theta \end{bmatrix}$, $\theta$ is the fast-axis angle of the QWP.

Setting $E_0 = \begin{bmatrix} A_0 \\ A_0 \end{bmatrix}$ and $\Delta x = 0$, using $E_+$ as the object of analysis, we can simplify Eq. (10) based on Eq. (5) as follows:

$$\eta_{+L} = \frac{A_+^2}{A_0^2} = \frac{1}{16}\eta_{max}\left[4K_1\sin^2(2\theta) + 2K_2\sin(2\theta)\sin(4\theta) + K_3\sin^2(2\theta)\right] \quad (12)$$

The coefficients $K_1$, $K_2$, $K_3$ and $k$ are related only to the grating diffraction parameters and are given by the following:

$$K_1 = k + 2\sqrt{k}\cos(\Delta\delta) + 1 \quad (13)$$

$$K_2 = -4\sqrt{k}\sin(\Delta\delta) \quad (14)$$

$$K_3 = k - 2\sqrt{k}\cos(\Delta\delta) + 1 \quad (15)$$

$$k = \frac{\eta_{min}}{\eta_{max}} \quad (16)$$

where $\eta_{max}$ and $\eta_{min}$ represent the larger and smaller values between $\eta_{TM}$ and $\eta_{TE}$, respectively, defining the relative energy conversion efficiency $\eta'_{+L}$ as:

$$\eta'_{+L} = \frac{\eta_{+L}}{\eta_{max}} \quad (17)$$

Eq. (12) and Eq. (17) characterize the relationship between the energy conversion efficiency $\eta_{+L}$ and the grating's diffraction characteristics, denoted as $\{k, \Delta\delta\}$. Our aim is to discern the parameter distribution for $\{k, \Delta\delta\}$ that maximizes energy conversion efficiency $\eta_{+L-max}$ and to understand how these diffraction parameters influence the distribution of peak efficiency, thereby establishing the specific criteria for optimal grating diffraction characteristics that maximize the efficiency of LCGI systems.

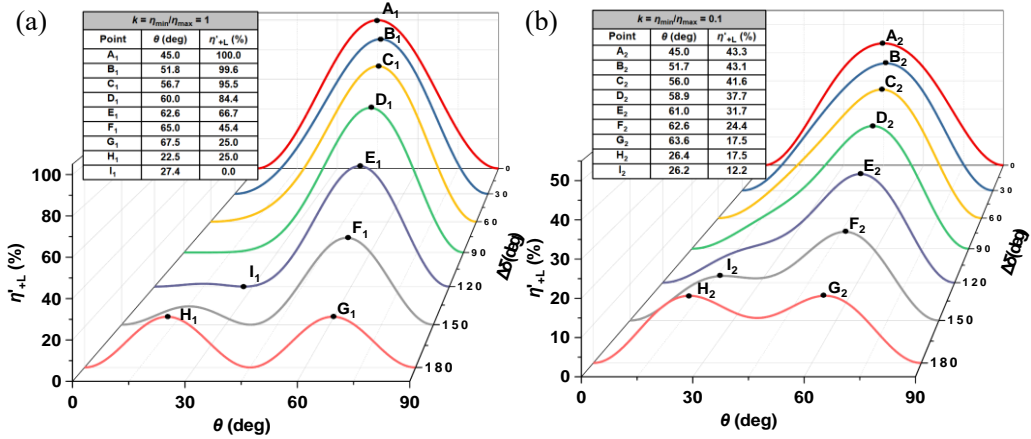

**Fig. 7.** Relative energy conversion efficiency of the system versus QWP fast-axis angle $\theta$ for various phase differences $\Delta\delta$, with marked extreme points (A-I). (a) $k = 1$, efficiency up to 100%. (b) $k = 0.1$, maximum efficiency is 43.3%.

Fig. 7, derived from Eq. (12) and Eq. (17), shows the relationship between the energy conversion efficiency $\eta'_{+L}$ and the QWP's fast-axis angle $\theta$, considering a phase difference and two $k$ values ($k = 1$ in (a) and $k = 0.1$ in (b)). The difference in $k$ results in a decrease in the amplitude of (a) and a shift in the peak value of the curve. Key findings include: maintaining equal diffraction efficiency in both TE and TM directions ($k = 1$) is crucial for enhancing the relative energy conversion efficiency; retaining the default circular polarization ($\theta = 45°$) in Littrow interferometers in the presence of a phase difference

leads to energy loss; and a large phase difference requires a broad initial selection of the waveplate's fast axis angle.

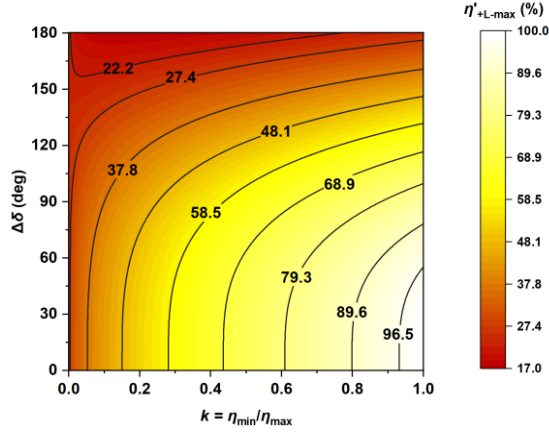

**Fig. 8.** Maximal relative energy conversion efficiency in LCGI with varied grating diffraction parameters $\{k, \Delta\delta\}$.

Fig. 8 displays the distribution of maximum relative energy conversion efficiency $\eta'_{+L\text{-max}}$ under varying grating diffraction characteristics $\{k, \Delta\delta\}$. It reveals a trend towards the grating's polarization-independent direction, peaking at 100% only at $k = 1$ and $\Delta\delta = 0$. Integrating conclusions from Fig. 7 and Eq. (17), we have identified the ideal grating criteria in LCGI: high diffraction efficiency, equal diffraction efficiency in TE and TM directions ($k \rightarrow 1$), and minimal phase difference ($\Delta\delta \rightarrow 0$). These criteria define the ideal fabrication direction for Cr grating.

Drawing from Fig. 7, evident that the use of QWP in LCGI can only achieve the maximum potential diffraction efficiency when $\Delta\delta = 0$. This prompts us to further explore the possibility of replacing the QWP in the LCGI. Due to space constraints, we briefly state that we can mathematically demonstrate that using a waveplate with different phase delays $\delta$ can eliminate the energy loss caused by $\Delta\delta$ and energy conservation is exhibited through Eq. (18):

$$\eta'_{+L\text{-max}} = \frac{1}{4}\left(\sqrt{k} + 1\right)^2 \tag{18}$$

Eq. (18) can also be written as:

$$\eta_{+L} = \frac{1}{4}\eta_{\max}\left(\sqrt{k} + 1\right)^2 = \frac{1}{4}\left(\sqrt{\eta_{TM}} + \sqrt{\eta_{TE}}\right)^2 \tag{19}$$

In this section, we analyzed the impact of grating diffraction characteristics $\{k, \Delta\delta\}$ (also understood as polarization sensitivity) on the energy conversion efficiency $\eta_{+L}$, thereby establishing the fabrication criteria of grating diffraction characteristics in LCGI. The next section will analyze the corresponding Cr grating groove.

### 2.3. Diffraction characteristics of Cr grating

In accordance with Eq. (8), the symmetric groove of Cr grating is a prerequisite for its optimal performance. In this section, we utilize the RCWA solver to delve into the impact of the Gaussian symmetric groove on the grating's diffraction characteristics.

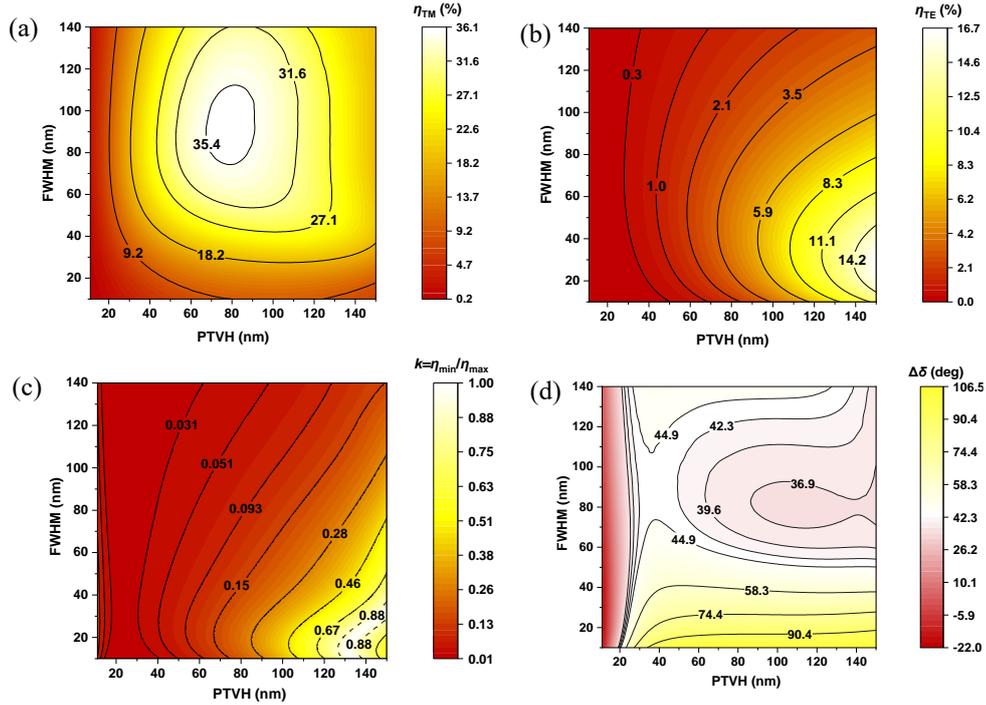

**Fig. 9.** Cr Grating's -1st Diffraction Efficiency and Phase Difference at LCGI Working Wavelength (405 nm) and Littrow Angle, with Cr Refractive Index $n$=2.04 - i 2.88[34]; (a) TM Direction, (b) TE Direction, (c) Efficiency Ratio $k$, (d) Phase Difference $\Delta\delta$ of Diffracted Light.

Fig. 9 provides a comprehensive overview of the diffraction characteristics of Cr grating. As illustrated in Fig. 9(a) and 9(b), the significant influence of PTVH and FWHM on the -1st diffraction efficiency is evident in both TM and TE modes, with TM consistently demonstrating higher values across a wide range. This inherent polarization sensitivity of Cr grating, as a subwavelength metal grating, is further showcased in Fig. 8(c), presenting the distribution of the efficiency ratio k, particularly in regions where $H > 125$ nm and $F < 40$ nm. Fig. 9(d) further supplements our understanding of this sensitivity by demonstrating the phase difference $\Delta\delta$ of the diffracted light. Most regions exhibit a phase shift of 30° ~ 40°, while negligible phase differences are observed for $H < 20$ nm. The significance of the $H$ parameter in Cr grating optimization is thus emphasized.

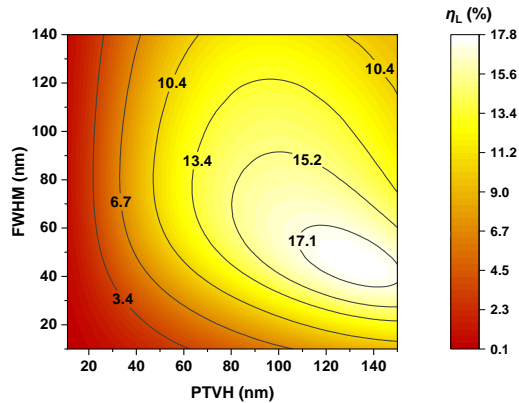

**Fig. 10.** Energy Conversion Efficiency Distribution of Cr Grating in a Self-Traceable Grating Interferometer.

Fig. 10 presents an analysis of the energy conversion efficiency of Cr grating within the self-traceable grating interferometer. The distribution diagram reveals a peak efficiency of 17.80% at $H = 132$ nm and $F = 46$ nm. However, when considering the stability of post-atom lithography experiments—where the FWHM often exceeds 45% of the PTVH [35]—the efficiency at the optimal efficiency drops to 14.95% at $H$=99 nm and $F$=96 nm after the second screening. Despite these variations, contour lines help identify the optimal groove. Our experience suggests that manipulating $H$ is simpler than $F$, but extremely low $H$ values increase Cr grating's line edge roughness, affecting the diffracted light's wavefront. Thus, we recommend prioritizing Cr grating fabrication with $H\geqslant 20$ nm to meet LCGI requirements. Regardless of $H$, employing the Channeling Atoms phenomenon during preparation is advised to reduce the FWHM of the grating groove and increase the energy conversion efficiency.

## 3. Results

### 3.1 Production process and characterization

Fig. 11 presents the process of creating an expanded Cr grating by our recently developed scanning atom lithography[19]. By moving the Dove prism vertically to scan the standing wave field ($\lambda$=425.55 nm), we facilitated a precise and wide-ranging deposition of Cr atoms at the wave nodes. Consequently, this method enabled us to expand the scale of the original Cr grating from the μm-scale to mm-scale (2mm×2mm). Notably, this method provides a grating with a pitch of $d=\lambda/2$=212.78 nm, virtually devoid of pitch value errors.

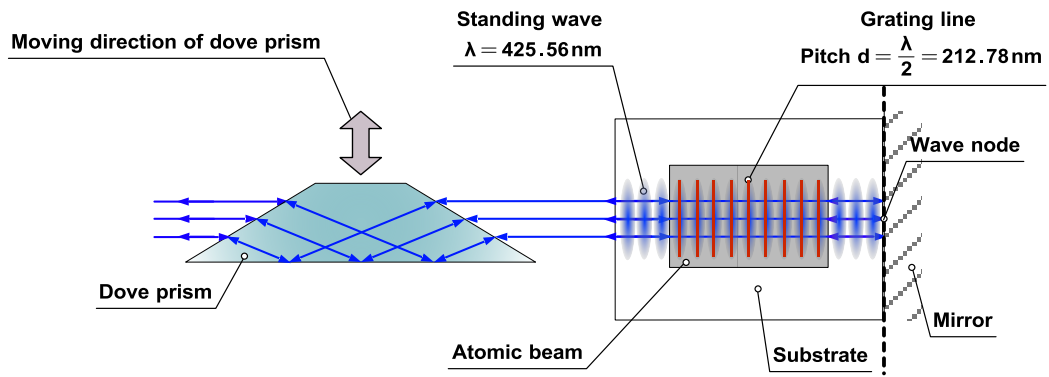

**Fig. 11.** Scanning Atom Lithography Schematic for Cr Grating Fabrication by Dove Prism Movement, Resulting in Virtually Error-Free Grating's pitch value.

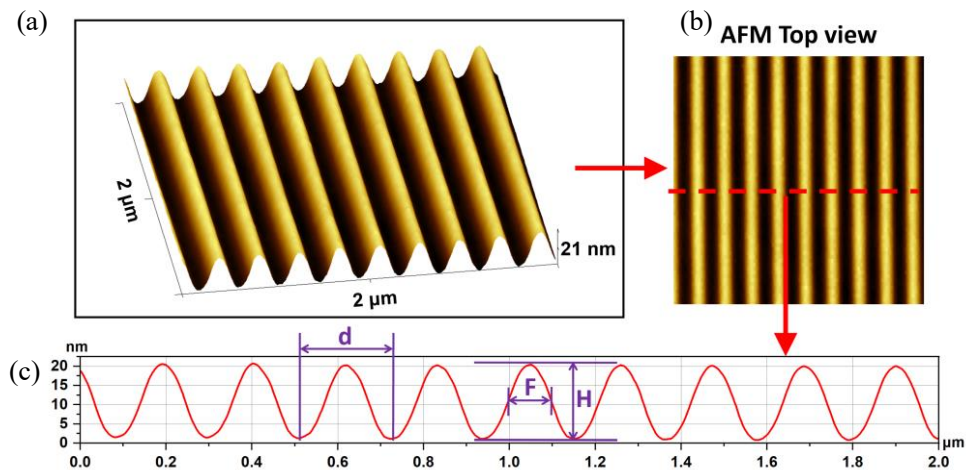

**Fig. 12.** AFM Characterization of the Fabricated Cr Grating

Fig. 12 presents the AFM characterization results of the Cr grating sample. Fig. 12(a) and (b) respectively depict the 3D and top-down AFM images, showing the nm-scale line edge roughness even at 4700 lines/mm. This vital attribute for nanometer-length metrology reduces wavefront distortion, enhancing grating interferometer precision. Fig. 12(c) displays a cross-sectional view of the Cr grating, showcasing Gaussian characteristics that align with the grating slot traits described by Eq. (1), thus resonating with our theoretical anticipations. The groove parameters were measured as follows: $d = 212.78$ nm, $H = 20.78$ nm and $F = 110.71$ nm. These results affirm the effectiveness of our Cr grating fabrication process, laying the groundwork for further experimental testing.

*3.2 Experimental testing*

Following the AFM characterization of the Cr grating, we proceeded with a series of tests focusing on its diffraction characteristics. This includes the integration of the fabricated Cr grating into a self-traceable grating interferometer system.

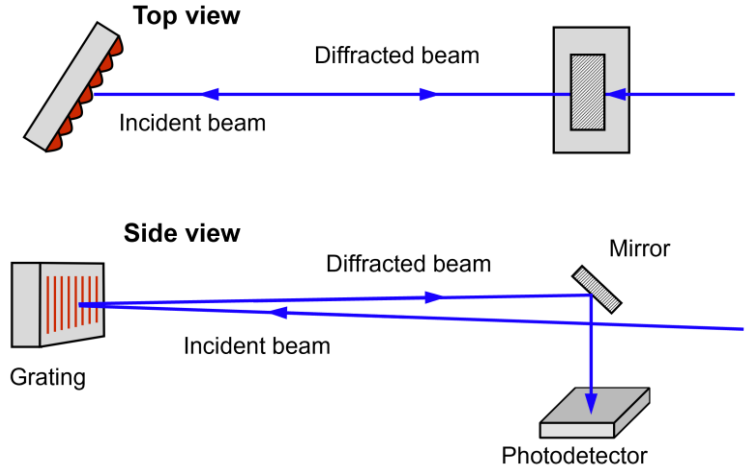

Fig. 13. Schematic of the Cr grating diffraction efficiency measurement at the Littrow angle.

Table.1 Diffraction efficiency measurement results of Cr grating.

|  | -1st $\eta_{TM}$ (%) | -1st $\eta_{TE}$ (%) | $\eta_L$ (%) |
|---|---|---|---|
| Left incident light | 5.12 | 0.45 | 2.02 |
| Right incident light | 5.17 | 0.47 | 2.07 |

Fig. 13 depicts the method for measuring the -1st order diffraction efficiency of the Cr grating. We employed a 405 nm light source and slightly deviated the grating from the Littrow incidence state. This approach still provides a close approximation to the true diffraction efficiency value[36].Tab. 1 shows the measured results under symmetric incidence. The close match in the diffraction efficiencies of the -1st order diffracted light validates the successful fabrication of a symmetric grating, while the less-than-maximum $\eta_L$ value (indicated by Eq. (19)) suggests the existence of a phase difference $\Delta\delta$ in the diffracted light. The discrepancies between Tab. 1 data and Section 3 simulation results may stem from various factors.(a) peak-valley height inconsistencies in the Cr grating produced by scanning atom lithography; (b) mismatch between the set and actual refractive index of Cr in the simulation; (c) loss caused by imperfections of optical components, among others. It is noteworthy that $\eta_{TM}$ significantly exceeds $\eta_{TE}$, thus failing to achieve the $k \rightarrow 1$ fabrication criteria. Given the geometric control limitations of the Gaussian groove shape (only $H$ and $F$ available) and the inherent constraints of Cr atomic surface growth broadening phenomena making it challenging to further reduce $F$, fabricating a polarization-independent Cr grating still remains a challenge. A potential solution could involve using a shorter UV light source to replace the 405 nm, but this would also limit the application scenarios of the self-traceable grating interferometer.

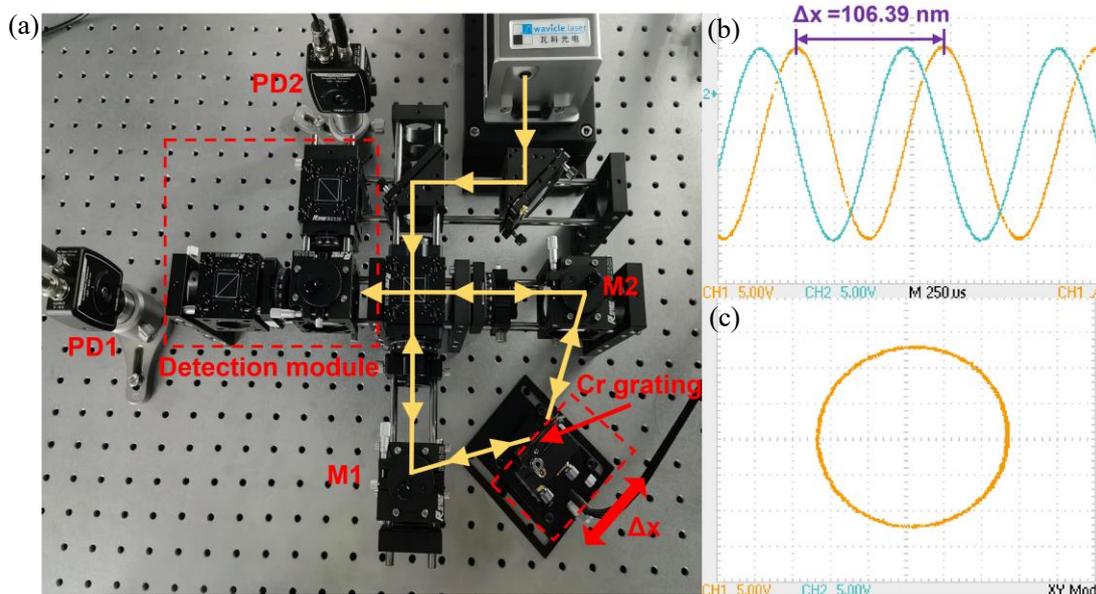

**Fig. 14.** Self-Traceable Grating Interferometer Based on Cr Grating. (a) Experimental setup with RayCage (Zhenjiang) cage structure, (b) Original interference signal, (c) Uncorrected Lissajous figures.

Fig. 14 showcases the configuration of the self-traceable grating interferometer and the interference signal obtained by driving a nano-motion table. Each sine cycle displayed in Fig. 14(b) corresponds to half of the Cr grating pitch ($d/2$=106.39nm), which is traceable to the Cr transition frequency $^7S_3 \rightarrow {}^7P_4^0$. This unique feature makes it unnecessary to pre-calibrate the grating pitch using laser diffraction or metrological AFM methods, substantially shortening the traceability path from the grating interferometer to the International System of Units (SI) [20]. Fig. 14(c) portrays the orthogonal signals in X-Y mode, yielding a Lissajous figure characterized by a minor DC bias and an almost circular shape. Compared to Si-lattice X-ray interferometers that also use natural constants as their measurement standards[11], our system shows a more favorable performance in the acquisition of interference signals. Consequently, we conclude that the fabricated Cr grating fulfills the optimization objectives required for interferometers, enabling its further application in length traceability chain measurements.

## 4. Conclusion

This research carried out optical design on the geometry and diffraction characteristics of Cr grating, which serves as the core self-traceable length standard in the new traceability chain. Firstly, we analyzed the bidirectional Gaussian properties of the Cr grating's geometry and obtained its diffraction characteristic distribution. Then, demonstrated the necessity of polarization-insensitive characteristics for optimal performance in Littrow configuration grating interferometers and established the experimental preparation direction for Cr grating. During the experiments, we fabricated an expanded Cr grating by scanning atom lithography, characterized its morphological features, measured its diffraction characteristics and performed preliminary verification in the self-traceable grating interferometer. This research aligns with the trend in metrology of directly reproducing natural constants following the reform of the International System of Units (SI), providing a significant reference for the development of new metrological technologies in the traceability chain. However, due to the limitations of the atom lithography experimental mechanism, we were unable to successfully fabricate a polarization-insensitive Cr grating, and the fabricated sample still exhibited uneven PTVH, indicating areas for further exploration and improvement in future research.